\documentclass{npw}
\usepackage{graphicx,colordvi}
\usepackage{amssymb}

\newcommand{\req}[1]{(\ref{#1})}
\newcommand{\bel}[1]{\begin{equation}\label{#1}}
\newcommand{\belar}[1]{\begin{eqnarray}\label{#1}}

\begin{document}
\title{The shell effects in the scission-point configuration   \\ 
  of fissioning nuclei}
\author{  F. A. Ivanyuk\email{ivanyuk@kinr.kiev.ua} \\
  \it Institute for Nuclear Research, Prospect Nauki 47, 03680 Kiev, Ukraine }
\pacs{02.60.Lj, 02.70.Bf, 21.60.Cs, 21.60.Ev, 25.85.Ec}
\date{\today}
\maketitle

\begin{abstract}

In the present work the formal definition of the scission point - the maximal 
elongation at which the nucleus splits into two fragments - is given. The 
shape and the deformation energy  at the scission point are calculated within the 
macroscopic-microscopic model.

Three minima in the scission point deformation energy are found corresponding to the 
"standard", "supershort" and "superlong" fission modes. The contribution of 
each fission mode to the mass distribution of the fission fragments and total 
kinetic energy is discussed and compared with the experimental results. On 
the example of the fission process of U-235 by thermal neutrons  it is shown that the 
present approach reproduces correctly the position of the peaks of the 
mass distribution of the fission fragments, the value and the fine details  of the total kinetic energy distribution and the magnitude of the total excitation energy of the fission fragments.

\end{abstract}

\section{Introduction}

In the theory of nuclear fission quasistatic quantities 
like the potential energy surface, the ground state energy and deformation, the fission barrier height, etc.,  
are commonly calculated within the macroscopic-microscopic method \cite{Str66,brdapa}. In this method the total energy of the fissioning nucleus consists of the two parts, macroscopic and microscopic. Both parts are calculated at the fixed shape of nuclear surface. In the past a lot of shape parameterizations were proposed and used. 
% \cite{brdapa,cohswi,koonin,pash71,marhun,nix,pombar}. 
A good choice of the shape parametrization is often a key to the success of the theory. 
Usually, one relies on physical intuition for the choice of the shape parametrization.

A method to define the shape of the nuclear surface which does not rely on any shape parametrization was proposed by V. Strutinsky in \cite{stlapo,strjetp45}. In this approach the shape of an axial, left-right symmetric nucleus was defined by looking for the minimum of the liquid-drop energy under the additional restrictions that fix the volume and elongation of the  drop.  

Recently the method was further developed \cite{fivan08,ivapom2} by incorporating the axial \cite{ivpoba,ivapom4} and left-right asymmetry \cite{procedia} of the nuclear shape. 

The important result of the Strutinsky procedure \cite{stlapo} is the possibility to definite in a formal way the scission point as the maximal elongation at which the nucleus splits into two fragments. 

In the present work the attention is focused on the scission point configuration. The shape at the scission point and the corresponding deformation energy are examined in detail 
and an approximation is suggested for the evaluation of the fission observables: the mass, the total kinetic energy and the total excitation energy distribution of the fission fragments. A similar investigation was carried out in \cite{CIP} where the shape of nucleus around the scission point was parametrized in terms of Cassini ovals.

The paper is organized as follows. Section \ref{optimal} contains a
short overview of the Strutinsky prescription \cite{stlapo}. 
Mass-asymmetric shapes are considered in detail in Sect. \ref{asymm}. The
scission point configuration and the deformation energy are defined in Sect.
\ref{scission}. In Sect.\ref{observa} the 
calculated results for the mass distribution of fission fragments, the total kinetic and excitation energies are compared with the experimental data for the fission of $^{235}U$ by thermal neutrons.
A short summary is given in Sect. \ref{summa}.
%%%%%%%%%%%%%%%%%%%%%%%%%%%%%%%%%%%%%%%%%%%%%%%%%%%%%%%%%%%%%%%%%%%%%%%%%%%%%%%%%%%%%%
\section{The optimal shapes of fissioning nuclei}
\label{optimal}
%%%%%%%%%%%%%%%%%%%%%%%%%%%%%%%%%%%%%%%%%%%%%%%%%%%%%%%%%%%%%%%%%%%%%%%%%%%%%%%%%%%%%%
It was suggested in \cite{stlapo} to describe the shape of a left-right and axial symmetric nucleus by some profile function $\rho(z)$. The shape of the surface is obtained then by rotation  of the $\rho(z)$ curve around the $z$-axis. 
A formal definition of $\rho(z)$ is obtained by searching
for the minimum of the liquid-drop energy, $E_{\rm
LD}=E_{\rm surf} + E_{\rm Coul}$, under the constraints that the
volume $V$ and the elongation $R_{12}$ are fixed,
 \bel{variation}
 \frac{\delta}{\delta \rho}\left[E_{\rm LD} - \lambda_1 V -\lambda_2 R_{12} \frac{}{}\! \right] = 0 ,
\end{equation}
with
 \bel{r12}
 R_{12}=\frac{2\pi}{V}\int\limits_{z_1}^{z_2}\rho^2(z)\vert z\vert dz \,\,,
\quad V=\pi\int\limits_{z_1}^{z_2}\rho^2(z) dz \,\,.
\end{equation}
In \req{variation} $\lambda_1$ and $\lambda_2$ are the
corresponding Lagrange multipliers. The elongation parameter
$R_{12}$ was chosen in \cite{stlapo} as the distance between
the centers of mass of the left and right parts of the nucleus.

The variation in \req{variation}
results in an
integro-differential equation for $\rho(z)$
 \bel{diffeq}
\rho\rho^{\prime\prime} \!= 1+(\rho^{\prime})^2 - \rho [ \lambda_1 +
\lambda_2 \vert z
 \vert
       - 10x_{\rm LD}\Phi_S ] \! \left[1+(\rho^{\prime})^2 \right]^\frac{3}{2}.
\end{equation}
Here $\Phi_S\equiv\Phi(z, \rho(z))$ is the Coulomb potential at
the nuclear surface, and $x^{}_{\rm LD}$ is the fissility parameter
of the liquid drop.

By solving Eq. (\ref{diffeq}) for given $x_{LD}$ and $\lambda_2$ ($\lambda_1$ is
fixed by the volume conservation condition) one obtains the profile function
$\rho(z)$ which we refer to as the {\it optimal shape}.

\begin{figure}[ht]
\includegraphics[width=\columnwidth]{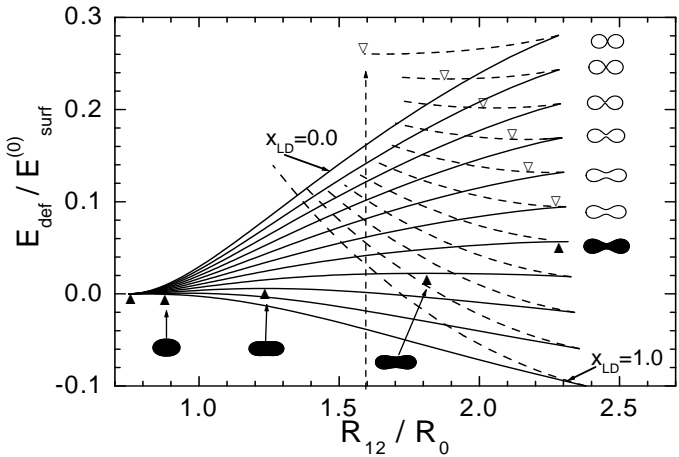}
\caption{\label{edef} Liquid-drop deformation energy
(\protect\ref{eldm}) as function of parameter $R_{12}$ for a few
values of the fissility parameter $x_{\rm LD}$.}
\end{figure}

The liquid drop deformation energy $E_{\rm def}^{LD}=E_{\rm LD}-E_{\rm LD}^{(0)}$
(in units  of the surface energy for a spherical shape)
 \bel{eldm}
E_{\rm def}\equiv E_{\rm def}^{LD}/{E_{\rm surf}^{(0)}}=B_{\rm surf}-1 + 2x_{\rm
LD}(B_{\rm Coul}-1)\,,
\end{equation}
calculated with these shapes
is presented in Fig.~\ref{edef}. In (\ref{eldm}) $B_{\rm Coul}\equiv {E_{\rm
Coul}}/{E_{\rm Coul}^{(0)}}$ and   $B_{\rm surf}\equiv {E_{\rm
surf}}/{E_{\rm surf}^{(0)}}$
where an index $^{(0)}$ refers to the spherical shape.
%%%%%%%%%%%%%%%%%%%%%%%%%%%%%%%%%%%%%%%%%%%%%%%%%%%%%%%%%%%%%%%%%%%%%%%%%%%%%%%%%%%%
%%%%%%%%%%%%%%%%%%%%%%%%%%%%%%%%%%%%%%%%%%%%%%%%%%%%%%%%%%%%%%%%%%%%%%%%%%%%%%%%%%%%%%%%

One can see from Fig.~\ref{edef} that the elongation $R_{12}$ of
the shapes shown in these figures is limited by some critical
value $R_{12}^{(crit)}$. Above this deformation mono-nuclear
shapes do not exist.  This critical deformation was interpreted in
\cite{stlapo} as the scission point. 
%%%%%%%%%%%%%%%%%%%%%%%%%%%%%%%%%%%%%%%%%%%%%%%%%%%%%%%%%%%%%%%%%%%%%%%%%%%%%%%%%%%%%%%
\section{The mass-asymmetric scission shapes}
\label{asymm}
%%%%%%%%%%%%%%%%%%%%%%%%%%%%%%%%%%%%%%%%%%%%%%%%%%%%%%%%%%%%%%%%%%%%%%%%%%%%%%%%%%%%%%%
The optimal-shape approach of \cite{stlapo} can be generalized to
mass-asymmetric shapes. For this goal one has to include into
Eq. \req{variation} one more constraint fixing the mass asymmetry
of the drop. 

Here we should note that the formal definition of the mass asymmetry does not exist. In the theory of fission (and fission experiments) one usually understand the mass asymmetry $\delta$ as the relative difference between left and right parts of the nucleus,
\bel{delta}
 \delta\equiv\frac{M_L-M_R}{M_L+M_R}=\frac{\pi}{V}\int \rm{Sign}(z-z^*)\rho^2(z) dz\,.
\end{equation}
In the case that the nuclear shape has a neck,
$z^*$ coincides with the position of the neck, $z^*=z_{neck}$. By $z_{neck}$ we mean here the point where the derivative $d\rho(z)/dz$ turns into zero.

This definition is meaningful for separated fragments or for 
shapes with a neck, i.e. at large deformation $R_{12}$. 
Shapes of small deformation, or 
pear-like shapes are commonly characterized by the quadrupole $Q_2$ and octupole $Q_3$ moments. In principle, both pairs of constraints $\{R_{12}, \delta\}$ or $\{Q_2, Q_3\}$ can be used for the definition of mass-asymmetric optimal shapes. 

However, it turns out \cite{procedia} that these two sets of constraints lead to rather different scission point shapes and deformation energies. For 
an accurate description of the mass-asymmetric nuclear shape
at the scission point it is necessary to construct an interpolation between the two sets of
definitions 
for the elongation and mass asymmetry which are applied successfully at small deformations  or for
separated fragments. Such an interpolation can be achieved by the replacement of the $\vert z \vert$ and Sign z
functions which appear in \req{diffeq} and \req{delta} by the following {\it smoothed} absolute value and Sign functions
\belar{smooth} \vert z\vert
\Longrightarrow f_2(z)\equiv\sqrt{z^2 + (\Delta z)^2}\,,\\
\rm{Sign}\,z \Longrightarrow f_3(z)\equiv\frac{z}{\sqrt{z^2 +
(\Delta z)^2}}.\nonumber
\end{eqnarray}
The expressions \req{r12} and \req{delta} for the elongation and mass asymmetry change then into
\belar{rsmooth} 
R_{12}\Longrightarrow\widetilde R_{12}\equiv\frac{\pi}{V}\int\limits_{z_1}^{z_2}f_2(z-z^*)\rho^2(z) dz \,\,,\\
\delta \Longrightarrow\widetilde \delta\equiv\frac{\pi}{V}\int\limits_{z_1}^{z_2}f_3(z-z^*)\rho^2(z) dz.\nonumber
\end{eqnarray}

The replacement \req{smooth} contains an additional parameter -
the smoothing width $\Delta z$. In principle, one can consider it
as an additional collective parameter which has to be taken into
account in the dynamical calculations. 
In the quasi-static limit
one could expect that the value of $\Delta z$ is close to the
neck radius in the scission region. In the calculations reported below $\Delta z$ was chosen 
as in \cite{procedia} as $\Delta z=0.25 R_0$. The calculated results do not change much if $\Delta z$ varies within the limits $0.1 R_0\lesssim \Delta z\lesssim 0.5 R_0$.
%%%%%%%%%%%%%%%%%%%%%%%%%%%%%%%%%%%%%%%%%%%%%%%%%%%%%%%%%%%%%%%%%%%%%%%%%%%%%%%%%%%%
\section{Scission shape and potential energy}
\label{scission}
\begin{figure}[h]
\includegraphics[width=\columnwidth]{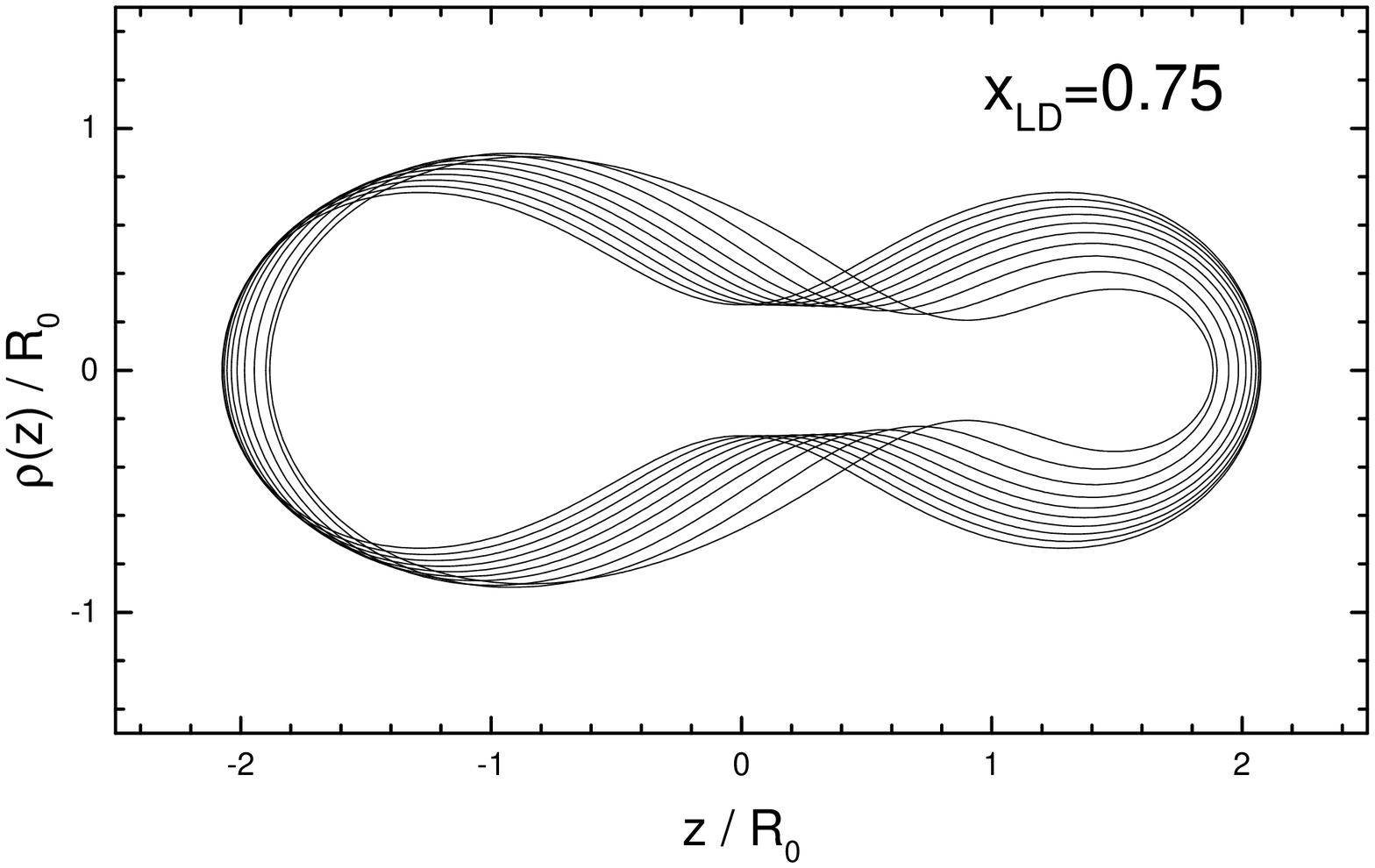}
\caption{\label{smoothed} Solutions of Eq.\ (\ref{diffeq2}) (with $\Delta z\!=\!0.25
R_0$) at the maximal elongation $R_{12}^{(crit)}$ for different values of the mass asymmetry $\delta=0, 0.1, ... 0.6$.}
\end{figure}

%%%%%%%%%%%%%%%%%%%%%%%%%%%%%%%%%%%%%%%%%%%%%%%%%%%%%%%%%%%%%%%%%%%%%%%%%%%%%%%%%%%%
The use of the constraints given in \req{rsmooth} leads to the following Euler-Lagrange equation 
\bel{diffeq2}
  \frac{\delta}{\delta \rho} \left[E_{\rm LD} - \lambda_1 V -\lambda_2  \widetilde R_{12}-\lambda_3\widetilde\delta \frac{}{}\right] = 0  \; .
\end{equation}
Eq.\req{diffeq2} can be solved in the same way as
Eq.\req{diffeq}. Some examples of the shapes at the scission point $R_{12}^{(crit)}$ (maximal possible value of $R_{12}$ at fixed $\lambda_3$) for a few values of the mass asymmetry $\delta$ are shown in Fig.~\ref{smoothed}.

The optimal shapes given by the solutions of \req{diffeq2} have 
two degrees of freedom - elongation and 
mass asymmetry. The neck radius for given elongation and mass asymmetry 
takes on the most favored value which corresponds to the minimum of the potential
energy. In dynamical calculations of the fission
process the neck radius is often considered as an independent
collective variable which can deviate from the one corresponding
to the bottom of the potential energy surface. 
%%%%%%%%%%%%%%%%%%%%%%%%%%%%%%%%%%%%%%%%%%%%%%%%%%%%%%%%%%%%%%%%%%%%%%%%%%%%%%%%%%%%
%%%%%%%%%%%%%%%%%%%%%%%%%%%%%%%%%%%%%%%%%%%%%%%%%%%%%%%%%%%%%%%%%%%%%%%%%%%%%%%%%%%%%%%% 

In order to include the neck degree of freedom in the optimal-shapes procedure it was suggested in \cite{procedia} to add in Eq. \req{diffeq2} the constraint $\lambda_4 f_4(z)$, where $f_4(z)$ is a measure of the amount of matter in the neck region,
\bel{f4} f_4=\frac{1}{V}\int dV
\rho^2(z)\exp{\left[-\left(\frac{z-z_{neck}}{\Delta z}\right)^2\right]}\,.
\end{equation}
For simplicity it was assumed that $\Delta z$ has the same meaning and value as in Eq. \req{smooth}. Such type of a constraint was introduced already in \cite{pom2002}.

The inclusion
of a constraint $\lambda_4 f_4(z)$ allows to vary the critical elongation.  However it turned out not to be enough. 
The shape of fissioning nuclei is defined not only by the liquid drop part of the deformation energy but also by the shell effects. It was shown in \cite{pash71} that due to the shell structure in the deformation energy of actinide nuclei there exists a deep minimum corresponding to the shape at which one part is almost spherical due to the large shell correction of the double magic fragment $^{132}Sn$ and another part is very elongated. Clearly, such a shape can not be obtained within the variational approach \req{diffeq2} where the deformation of the fragments is completely fixed by the deformation dependence of the liquid-drop energy. 

In order to account for the influence of shell effects on the shape of the scissioning nucleus, it is necessary to include into the optimal shapes more degrees of freedom. For this purpose, instead of $\lambda_4 f_4(z)$, we include in the variational procedure \req{diffeq2} another two constraints which allow to vary 
%separately 
the shape of left and right fragments,
\bel{diffeq3}
  \frac{\delta}{\delta \rho}[E_{\rm LD} - \lambda_1 V -\lambda_2\widetilde R_{12}-\lambda_3\widetilde\delta - \lambda_5 Q_{2L} - \lambda_6 Q_{2R}]= 0 \,.
\end{equation}
In \req{diffeq3} $Q_{2L}$ and $Q_{2R}$ are the quadrupole moments of 
left and right parts of nucleus where the limit between {\it left} and {\it right} is defined at the neck position $z_{neck}$ where $\rho^{\prime}(z)$ vanishes. 
In what follows we are interested only in 
shapes at the scission point where the neck position is well defined.  Solving \req{diffeq3} 
for large values of $\lambda_5$ and $\lambda_6$
one gets very elongated ($\lambda_5$ and $\lambda_6$ both large positive) or very short shapes ($\lambda_5$ and $\lambda_6$ both large negative) or 
shapes where one part is close to spherical and another very elongated (opposite sign of $\lambda_5$ and $\lambda_6$).                                                                   
%%%%%%%%%%%%%%%%%%%%%%%%%%%%%%%%%%%%%%%%%%%%%%%%%%%%%%%%%%%%%%%%%%%%%%%%%
\begin{figure}[h]
\includegraphics[width=\columnwidth]{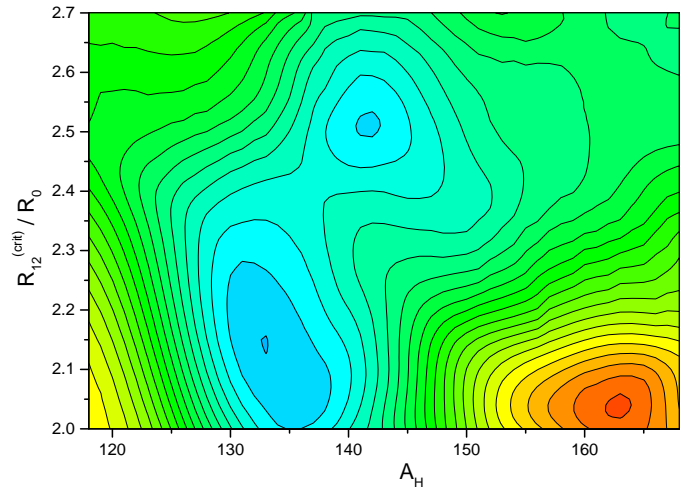}
\caption{\label{shco} The shell component of the scission point  deformation energy of $^{236}U$ as function of the heavy fragment mass number $A_H$ and the maximal elongation $R_{12}^{(crit)}$.}
\end{figure} 
%%%%%%%%%%%%%%%%%%%%%%%%%%%%%%%%%%%%%%%%%%%%%%%%%%%%%%%%%%%%%%%%%%%%%%%%%%%%%%%%%%%%

For each value of the mass asymmetry the scission point turns in this case into the ``scission surface''. In order to visualize this scission surface energy we reduce it to a  ``scission line''. Each point on the scission surface is characterized by the deformations of left and right fragment 
$Q_L$ and $Q_R$ and by the critical elongation $R_{12}^{(crit)}$ (which is a function of $Q_L$ and $Q_R$). For each value of $R_{12}^{(crit)}$ the {\it total deformation energy} (liquid drop part plus shell correction) was minimized with respect 
to the fragments deformation $Q_L$ and $Q_R$. In this way one gets the deformation energy at the scission point as function of the critical elongation $R_{12}^{(crit)}$. 
Figs.~\ref{shco} and \ref{eldm} show separately the shell correction and the liquid drop energy
obtained as the result of the above described minimization. 

One clearly notices two minima in Fig.~\ref{shco}. One, at 
{\it short} scission shape at $A_H\approx 134$ is caused by the large shell correction for $^{132}Sn$. The other minimum is caused by the ``deformed shell" at $A_H\approx 142$. As will become clear below, these minima have 
a large influence on the fission fragment mass distribution and the total kinetic energy. 

\begin{figure}[ht]
\includegraphics[width=\columnwidth]{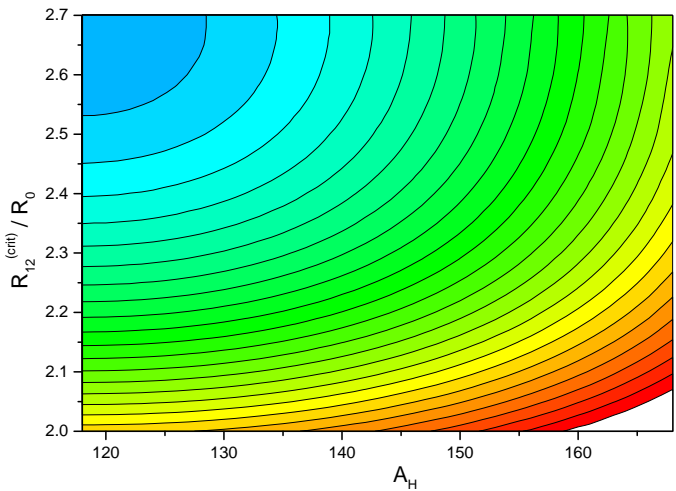}
\caption{\label{figeldm} The liquid drop component of the scission point  deformation energy of $^{236}U$ as function of the heavy fragment mass number $A_H$ and the maximal elongation $R_{12}^{(crit)}$.}
\end{figure}

As one could expect, the liquid-drop energy grows with 
increasing mass asymmetry and decreases with growing critical deformation $R_{12}^{(crit)}$
as seen on Fig.~\ref{figeldm}.
%%%%%%%%%%%%%%%%%%%%%%%%%%%%%%%%%%%%%%%%%%%%%%%%%%%%%%%%%%%%%%%%%%%%%%%%%%%%%%%%%%%%%%%%
\\[ -2.0ex]
%%%%%%%%%%%%%%%%%%%%%%%%%%%%%%%%%%%%%%%%%%%%%%%%%%%%%%%%%%%%%%%%%%%%%%%%%%%%%%%%%%%%%%%%

\begin{figure}[h]
\includegraphics[width=\columnwidth]{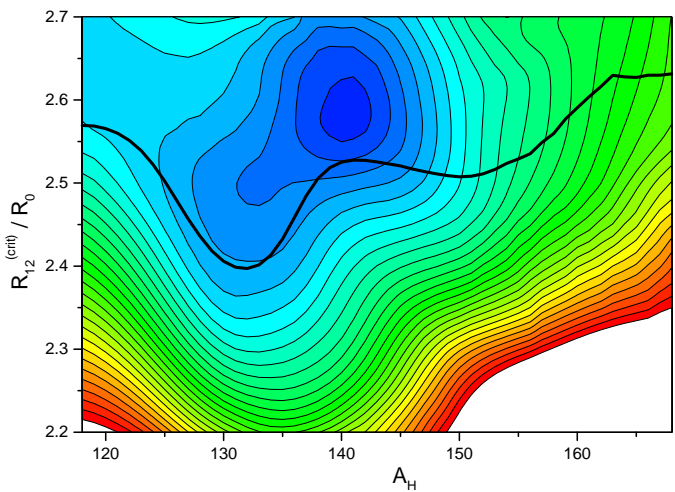}
\caption{\label{etot}
Total energy (liquid drop plus 
shell correction) for $^{236}U$ at the scission point as function of the heavy fragment mass number $A_H$ and the maximal elongation $R_{12}^{(crit)}$. The mean value \protect\req{rmean} of $R_{12}^{(crit)}$ is shown by thick line.}
\end{figure}

The total scission-point deformation energy for $^{236}U$ is shown in Fig.~\ref{etot}.
Due to overlay of the liquid drop and shell components of the energy, the 
minimum around $A_H=134$ is no longer seen. However, it influences the calculated results for the fission observables (see below).
%%%%%%%%%%%%%%%%%%%%%%%%%%%%%%%%%%%%%%%%%%%%%%%%%%%%%%%%%%%%%%%%%%%%%%%%%%%%%%%%%%%%
\\[ -5.0ex]                                                               
%%%%%%%%%%%%%%%%%%%%%%%%%%%%%%%%%%%%%%%%%%%%%%%%%%%%%%%%%%%%%%%%%%%%%%%%%%%%%%%%%%%%%%%%
%%%%%%%%%%%%%%%%%%%%%%%%%%%%%%%%%%%%%%%%%%%%%%%%%%%%%%%%%%%%%%%%%%%%%%%%%%%%%%%%%%%%%%%%
\section{Comparison with experimental data}
\label{observa}
%%%%%%%%%%%%%%%%%%%%%%%%%%%%%%%%%%%%%%%%%%%%%%%%%%%%%%%%%%%%%%%%%%%%%%%%%%%%%%%%%%%%%%%%
{\bf  The yield.} 
The description of the fission process requires a solution of the
dynamical problem. 
Unfortunately, the description of fission based on the Langevin equations for the time evolution of some shape variables \cite{abe} is very time consuming. 
Sometimes (see e.g.\ the recent work of Ref.\ \cite{cahaivta})
one describes the fission fragment distributions in terms of 
fission modes introduced by Brosa et al. \cite{brosa}. In this way one parametrizes the mass distribution of the fission fragments by a few Gaussians. The parameters of these  Gaussians are fixed {\it by a fit} to experimental results.

Having at ones disposal the potential energy surface shown in Figs.~\ref{shco}-\ref{etot}, it 
is worth to try {\it to calculate} the distributions measured in the %fission 
experiments, such as the fis\-sion-fragment mass distribution, the total excitation and kinetic-energy distribution and the neutron multiplicity.

Keeping in mind that fission is a {\it slow} process, one could assume
that during the fission process the state of the fissioning
nucleus is close to
thermal equilibrium, i.e.\ 
each point $q_i$ on the deformation-energy surface is
populated with a probability given by the Boltzmann distribution,
\bel{boltz} P(\delta_i, q_i)=e^{-\left(\frac{E(\delta_i, q_i)-F}{T_{coll}}\right)}\,,F\equiv -T_{coll}\log\sum_ie^{-\left(\frac{E(\delta_i, q_i)}{T_{coll}}\right)}\,.
\end{equation}
Here $T_{coll}$ is a parameter characterizing the width of the distribution \req{boltz} in the space of deformation parameters.
$E(\delta_i, q_i)$ in \req{boltz} is the sum of the liquid-drop deformation 
energy \req{eldm} and of the shell correction $E_{\rm shell}$, as shown in Fig.~\ref{etot}.

The distribution \req{boltz} is a basic assumption of the
scission-point model suggested in \cite{spm} (see also \cite{krapom}). 
The parameters of this model were fitted in \cite{spm} so as
to reproduce the numerous experimental data. 
$T_{coll}$ was found to be close to 1 MeV.
In the calculations shown below we used the  
value $T_{coll}$=1.5 MeV.

The normalized mass distribution of the fission fragments $Y(\delta)$
can be expressed then in terms of the deformation energy at the critical deformation
$R_{12}^{(crit)}$, 
\bel{yield}
Y(\delta)=\sum_i P(\delta, q_i)\,.
\end{equation}

In \req{boltz} the summation is carried out 
over the set $\{Q_L, Q_R, R_{12}^{(crit)}(Q_L, Q_R)\}$. 
Below we have calculated some distributions for the fission of $^{236}U$ for which 
experimental data are available. 
The fission of $^{235}U$ by thermal neutrons was analyzed in details in \cite{haru}. It was shown that the experimental fission-fragment mass distribution of \cite{humbsch} can be decomposed into three peaks centered around  $A_H=141$ (main peak), $A_H=134$ (three times lower) and a much smaller one ($~10^{-3}$ of the main peak) at the symmetric splitting.

\begin{figure}[ht]
\includegraphics[width=\columnwidth]{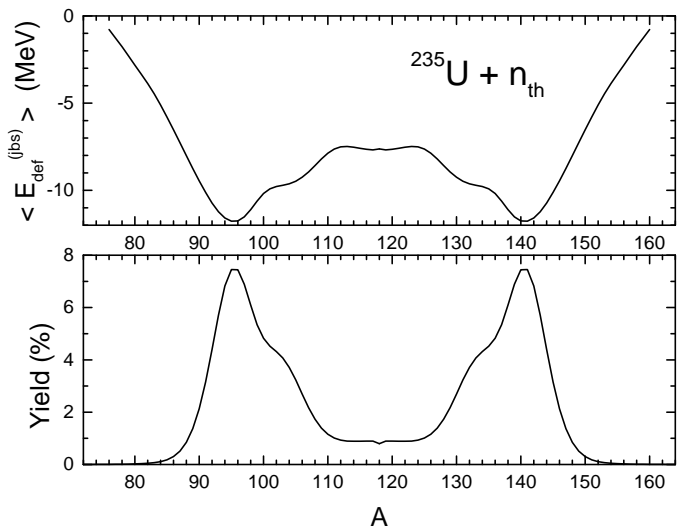}
\caption{\label{figyield} Top: 
Mean value \protect\req{emean} of the deformation energy  just before scission. 
Bottom: Calculated fragments yield \protect\req{yield}  for the fission of $^{235}U$ by thermal neutrons.}
\end{figure}

The top part of Fig.~\ref{figyield} 
shows the mean value of the deformation energy of $^{236}U$ at the scission point
\bel{emean}
\langle E_{def}(\delta)\rangle=\sum_i E_{def}(\delta, q_i)P(\delta, q_i)/\sum_i P(\delta, q_i)\,.
\end{equation}
whereas the bottom part
shows the yield \req{yield}. In principle, the contributions of all three minima of the deformation energy shown in Figs.~\ref{shco}, \ref{figeldm} are seen in the yield  \req{yield}. The ``superlong" minimum of the liquid-drop energy causes
the raise of the yield for symmetric splitting, the  ``standard" minimum at $A_H\approx 141$ is responsible for the main peak of the yield and the  ``supershort" minimum at $A_H\approx 134$ contributes to another peak of the mass distribution of fission fragments. 
%%%%%%%%%%%%%%%%%%%%%%%%%%%%%%%%%%%%%%%%%%%%%%%%%%%%%%%%%%%%%%%%%%%%%%%%%%%%%%%%%%%%
%%%%%%%%%%%%%%%%%%%%%%%%%%%%%%%%%%%%%%%%%%%%%%%%%%%%%%%%%%%%%%%%%%%%%%%%%%%%%%%%%%%%%%%%
The relative magnitude of these peaks deviates, however, substantially from the results of the analysis in \cite{haru}.

{\bf The total kinetic energy.} The measured distribution of the total kinetic energy is rather complicated. First of all one notices a sharp drop of the order of 10-15 MeV around symmetric fission. Such a drop is typical for the fission of all actinide nuclei. There is also a peak around $A_H=130$, the position of which does, however, not coincide with the main peak of the yield. 
Evidently
these features should be related somehow to the shell structure of the fissioning nucleus.

To calculate the kinetic energy distribution we note that it consists of two parts: the pre-scission kinetic energy and the Coulomb repulsion energy of the fragments immediately after scission. During the fission process, due to the friction force, part of the collective kinetic energy can turn into
%the 
heat and the rest would contribute to the pre-scission kinetic energy. Within the quasi-static limit we can give only an estimate for the pre-scission kinetic energy exploiting the fact that in case of fission of $^{235}U$ by thermal neutrons the excitation energy and the fission barrier height are approximately the same, so the excitation energy of $^{236}U$ at the barrier is very small. 
From the calculations of the friction tensor within the linear response theory \cite{ivahof} it is known that at small excitations the friction force is 
negligibly
small. So, in the case considered here, one could neglect the dissipative effects and estimate the pre-scission kinetic energy from the energy balance: the ground state energy $E_{gs}$ of $^{236}U$ plus the neutron binding energy $B_n$ is equal to the potential energy just before 
scission $E^{jbs}$ plus the pre-scission kinetic energy $KE_{pre}$, i.e.,
\bel{kepre}
KE_{pre}=E_{gs}+B_n-E^{jbs}\,.
\end{equation}
As $E^{jbs}$ we use here the mean value of the energy shown in the upper part of Fig.~ \ref{figyield}. The bump of $E^{jbs}$ for symmetric splitting leads then to 
a lowering of $KE_{pre}$ and, consequently, the total kinetic energy for symmetric splitting.

Another contribution to the total kinetic energy comes from the Coulomb repulsion energy $E_{Coul}^{int}$ immediately after 
scission. This quantity is defined mainly by the distance between centers of mass of the future fragments just before 
scission. The measured dependence of the total kinetic energy on the fragment mass can be explained if for symmetric splitting the main value of $R_{12}^{(crit)}$ 
\bel{rmean}
\langle R_{12}^{(crit)}(\delta)\rangle=\sum_i R_{12}^{(crit)}(\delta, q_i)P(\delta, q_i)/\sum_i P(\delta, q_i)\,
\end{equation}
would be relatively large and around $A_H=130$ the main value of $R_{12}^{(crit)}$ would be relatively small.
%%%%%%%%%%%%%%%%%%%%%%%%%%%%%%%%%%%%%%%%%%%%%%%%%%%%%%%%%%%%%%%%%%%%%%%%%%%%%%%%%%%%
\begin{figure}[b]
\includegraphics[width=\columnwidth]{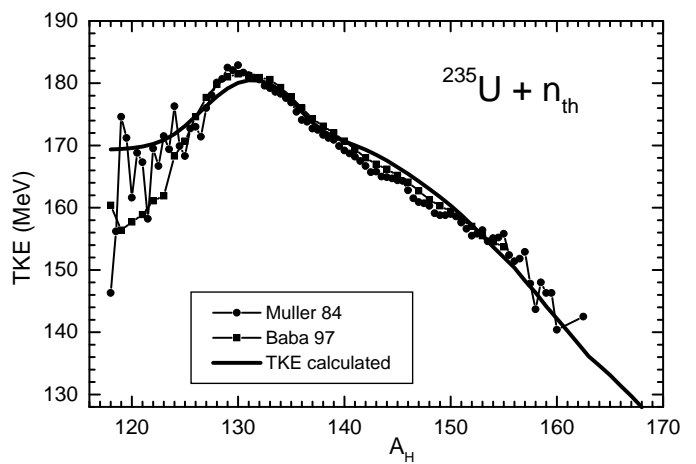}
\caption{\label{tke} The calculated (solid line) and the measured (circles \protect\cite{mul}, squares \protect\cite{baba}) total kinetic energy of fission fragments.}
\end{figure} 
%%%%%%%%%%%%%%%%%%%%%%%%%%%%%%%%%%%%%%%%%%%%%%%%%%%%%%%%%%%%%%%%%%%%%%%%%%%%%%%%%%%%%%%%

To calculate the Coulomb interaction energy it is assumed here that immediately after 
scission the shape of the system is given by the optimal shape of two separated fragments with 
a mass asymmetry $\delta$ placed at the distance $R_{12}^{(crit)}(\delta, Q_L, Q_R)$ between their centers of mass. As  
mentioned in \cite{procedia} the optimal shape of the fragments immediately after 
scission is very close to two spheres. 
Small quadrupole deformations of the fragments can be neglected. The Coulomb repulsion energy for such a configuration is just the repulsion energy of two point charges.
As one can see from Fig.~\ref{etot} the $\langle R_{12}^{(crit)}(\delta)\rangle$ (thick solid curve) has required maximum at $A=118$ and minimum around $A_H=132$, what is consistent with the experimental data. Together with the minimum of the pre-scission kinetic energy this leads to the minimum of the total kinetic energy at a symmetric splitting. 
As one can see from Fig.~\ref{tke} the calculated and the measured total kinetic energies are in a good agreement. 
This is at least an indication that the shape calculated just before scission is correct.

In these calculations the value of $r_0$ that appears in the Coulomb energy of a spherical nucleus was 
taken as $r_0=1.225\, \rm{fm}$. This value was fixed 
by Pashkevich \cite{pash71} through a
comparison of calculated and measured fission barrier heights. The same value was used here for the calculation of the optimal shapes. 

\begin{figure}[ht]
\includegraphics[width=\columnwidth]{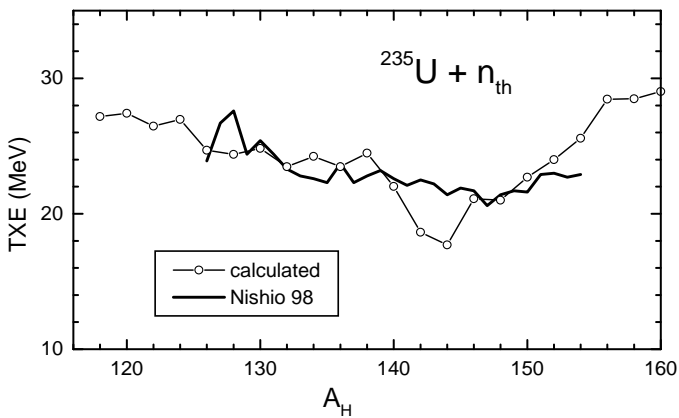}
\caption{\label{figtxe}
Measured \protect\cite{nishio98} and 
calculated (open circles)  total excitation energy \protect\req{txe} for the fission of $^{235}U$ by thermal neutrons.}
\end{figure}

{\bf The total excitation energy.} The total excitation energy $TXE$ is easy to calculate from the energy conservation condition
\bel{txe}
E_{gs}(A_L+A_H)+B_n=E_{gs}(A_L)+E_{gs}(A_H)+TKE+TXE\,.
\end{equation}
In \req{txe} 
$E_{gs}$ is the ground-state energy of the mother nucleus or the fragments. The ground-state energy is  calculated by the  macroscopic-microscopic method (or 
it can be taken from the published tables \cite{moeller95}). The total kinetic energy is calculated as explained above.
The resulting $TXE$ is compared with the experimental results in Fig.~\ref{figtxe}. The deviation between them is of the order of 10-20\% 
and the agreement is therefore only of qualitative nature. 
%%%%%%%%%%%%%%%%%%%%%%%%%%%%%%%%%%%%%%%%%%%%%%%%%%%%%%%%%%%%%%%%%%%%%%%%%%%%%%%%%%%%
%\\[ -7.0ex]
%%%%%%%%%%%%%%%%%%%%%%%%%%%%%%%%%%%%%%%%%%%%%%%%%%%%%%%%%%%%%%%%%%%%%%%%%%%%%%%%%%%%%%%%
\section{Summary}
\label{summa}
%%%%%%%%%%%%%%%%%%%%%%%%%%%%%%%%%%%%%%%%%%%%%%%%%%%%%%%%%%%%%%%%%%%%%%%%%%%%%%%%%%%%%%%%
In the present work a method to calculate the shape of fissioning nuclei at maximal deformation where the nucleus splits into two fragments is proposed. The method takes into account both the liquid drop properties and the shell structure of atomic nuclei. The deformation energy of the nucleus Uranium-236 just before scission is examined in detail and an 
approximation for the evaluation of the basic quantities measured in the fission  experiments such as the fission-fragment mass distribution, the total excitation and the kinetic energy of the fission fragments is proposed. It is demonstrated that the calculated distributions of fission fragments are in qualitative agreement with experimental data. 

\begin{ack}
The author would like to express his gratitude to Profs. J.~Bartel, N. Carjan and K. Pomorski for useful discussions and to the
Theoretical Physics Division, UMCS  for the hospitality during his stay at Lublin were the part of results reported in this paper was obtained. 
\end{ack}

\vspace{-5mm}
%%%%%%%%%%%%%%%%%%%%%%%%%%%%%%%%%%%%%%%%%%%%%%%%%%%%%%%%%%%%%%%%%%%%%%%%%%%%%%%%

\end{document}